\documentclass{aa}
\usepackage{natbib}
\usepackage{graphicx}
\usepackage{txfonts}
\usepackage[english]{babel}

\def\Msun{\hbox{$M_{\odot}$}}               
\def\Lsun{\hbox{$L_{\odot}$}}               
\def\Rstar{\hbox{$R_{\star}$}}              
\def\Tstar{\hbox{$T_{\star}$}}              
\def\Mdot{\hbox{$\dot{M}$}}               
\def\kms{\hbox{km$\,$s$^{-1}$}}            

\bibpunct{(}{)}{;}{a}{}{,}
\newcommand{\bftheta}{ {\mbox{\boldmath $\theta$}} }

\begin{document}

\title{The variable mass loss of the AGB star \object{WX Psc} as
  traced by the CO $J$\,=\,1$-$0 through 7$-$6 lines and the dust emission.}

\author{L.\ Decin \inst{1,2}\thanks{\emph{Postdoctoral Fellow of the
      Fund for Scientific Research, Flanders}} \and S.\ Hony \inst{3,1}
  \and A.\ de Koter \inst{2} \and G.\ Molenberghs \inst{4} \and
  S. Dehaes \inst{1} 
\and F. Markwick-Kemper\inst{5,6}
}
\offprints{L.\ Decin, e-mail: Leen.Decin@ster.kuleuven.ac.be}

\institute{
  Department of Physics and Astronomy, Institute for Astronomy,
  K.U.Leuven, Celestijnenlaan 200B, B-3001 Leuven, Belgium
  \and Sterrenkundig Instituut Anton Pannekoek, University of
  Amsterdam, Kruislaan 403 1098 Amsterdam, The Netherlands
  \and Service d'Astrophysique, CEA Saclay, F-91191 Gif-sur-Yvette,
  France
  \and Biostatistics, Center for Statistics, Universiteit Hasselt,
  Universitaire Campus, Building D, B-3590 Diepenbeek, Belgium
  \and School of Physics and Astronomy, University of Manchester,
  Sackville Street, PO Box 88, Manchester M60 1QD, UK 
  \and University of Virginia, Department of Astronomy, PO Box 400325,
  Charlottesville, VA 22904-4325, USA 
}

\date{received date; accepted date}


\abstract
{ Low and intermediate mass stars lose a significant fraction of
  their mass through a dust-driven wind during the Asymptotic Giant
  Branch (AGB) phase. Recent studies show that winds from late-type
  stars are far from being smooth. Mass-loss variations occur on
  different time scales, from years to tens of thousands of years.
  The variations appear to be particularly prominent towards the end
  of the AGB evolution. The occurrence, amplitude and time scale of
  these variations are still not well understood.
}
{The goal of our study is to gain insight into the structure of the
  circumstellar envelope (CSE) of \object{WX Psc} and map the possible
  variability of the late-AGB mass-loss phenomenon. }
{We have performed an in-depth analysis of the extreme infrared
  AGB star \object{WX Psc} by modeling \emph{(1)} the CO $J$\,=\, 1$-$0
  through 7$-$6 rotational line profiles \emph{and} the full spectral
  energy distribution (SED) ranging from 0.7 to 1300\,$\mu$m. We hence
  are able to trace a geometrically extended region of the CSE.}
{Both mass-loss diagnostics bear evidence of the occurrence of
  \emph{mass-loss modulations} during the last $\sim$2000\,yr. In particular,
  \object{WX Psc} went through a high mass-loss phase (\Mdot $\sim 5
  \, 10^{-5}$\,\Msun/yr) some 800\,yr ago. This phase lasted about
  $600$\,yr and was followed by a long period of low mass loss (\Mdot
  $\sim 5 \, 10^{-8}$\,\Msun/yr). The present day mass-loss rate is
  estimated to be $\sim 6 \, 10^{-6}$\,\Msun/yr.}
{The AGB star \object{WX Psc} has undergone strong mass-loss rate
  variability on a time scale of several hundred years during the last
  few thousand years. These variations are traced in the strength and
  profile of the CO rotational lines and in the SED. We have
  consistently simulated the behaviour of both tracers using radiative
  transfer codes that allow for non-constant mass-loss rates.}

\authorrunning{L.\ Decin et al.}
\titlerunning{The variable mass-loss of the AGB star \object{WX Psc}}

\maketitle


\section{Introduction}

The interstellar medium (ISM) is slowly enriched in heavy elements
through material ejected by evolved stars. Such stars lose mass
through a stellar wind, which is slow and dusty for cool giants and
supergiants, or through supernova explosions. The ejected material
merges with the interstellar matter and is later incorporated into new
generations of stars and planets (the so-called `cosmic cycle'). The
gas and solid dust particles produced by evolved stars play a major
role in the chemistry and energy balance of the ISM, and in the
process of star formation.

\vspace*{1.5ex} Low and intermediate mass stars (M $\sim 0.6 -
8$\,\Msun) are particularly important in this process since their mass
loss dominates the total ISM mass. About half of the heavy elements,
recycled by stars, originate from stars of this mass interval
\citep{Maeder1992A&A...264..105M}. Moreover, these stars are the
dominant source of dust in the ISM. When these stars leave the main
sequence, they ascend the red giant and asymptotic giant branches (RGB
and AGB). During these evolutionary phases, very large amplitude and
long period pulsations \citep[e.g.][]{Bowen1988ApJ...329..299B} lift
photospheric material to great heights. In these cool and dense
molecular layers dust grains condense.  Radiation pressure efficiently
accelerates these dust grains outwards and the gas is dragged
along. The resulting mass-loss rate is much larger than the
rate at which the material is burned in the core and dominates the
stellar evolution during this phase.  Despite the importance of the
mass-loss process in terminating the AGB stellar evolutionary phase
and in replenishing the ISM with newly-produced elements, the nature
of the wind driving mechanism is still not well understood
\citep[e.g.][]{Woitke2006A&A...460L...9W}.

Recent CO and scattered light observations of AGB objects show
that winds from late-type stars are far from being smooth
\citep[e.g.][]{Mauron2006A&A...452..257M}. Density variations in the
circumstellar envelope (CSE) occur on different time scales, ranging
from years to tens of thousands of years. Stellar pulsations may cause
density oscillations on a time scale of a few hundred days. A nuclear
thermal pulse may be responsible for variations every ten
thousand to hundred thousand years. Oscillations on a time scale of
a few hundred years are possibly linked to the hydrodynamical properties
of the CSE (see also Sect.~\ref{variability}).

The mass-loss variations appear to be particularly more prominent when
these stars ascend the AGB and become more luminous. It is often
postulated and in rare cases observed
\citep[e.g.][]{Justtanont1996ApJ...456..337J,
vanLoon2003MNRAS.341.1205V} that the AGB evolution ends in a very high
mass-loss phase, the so-called superwind phase
\citep{Iben1983ARA&A..21..271I}. The superwind mass-loss determines
the quick ejection of the whole envelope and the termination of the
AGB phase.

To enlarge our insight in this superwind mass-loss phase, and in
possible mass-loss variations occurring during this phase, we focus
this research paper on the study of \object{WX Psc}
(=\object{IRC+10\,011}, \object{IRAS 01037+1219}, \object{CIT 3}), one
of the most extreme infrared (IR) AGB objects known
\citep{Ulrich1966ApJ...146..288U}.  \object{WX Psc} belongs to the
very late-type AGB stars having a spectral type of M9\,--\,10
\citep{Lockwood1985ApJS...58..167L}. Given this spectral type and its
very red color ($V-K \ga 18$), its effective temperature can be
estimated to be $\la 2500$\,K \citep{Hofmann2001AA...379..529H}. \object{WX Psc}
is an oxygen-rich, long-period variable \citep[P
${\sim}$660\,days; ][]{LeBertre1993A&AS...97..729L}.

The Infrared Space Observatory (ISO) data show that \object{WX Psc} is
surrounded by an optically thick dusty shell, of which the presence of
crystalline silicate features \citep{Suh2002MNRAS.332..513S} is
indicative of a high mass-loss rate \citep{kemper01}. In the
circumstellar envelope (CSE) SiO \citep{Desmurs2000A&A...360..189D},
H$_2$O \citep{Bowers1984ApJ...285..637B} and OH
\citep{Olnon1980A&AS...42..119O} maser emission lines are detected
with expansion velocities ranging from 18 to 23\,km\,s$^{-1}$.

Estimates of the mass-loss rate, based on the modeling of CO
rotational line intensities or (near)-infrared (dust) observations,
span a wide range. Derived values lie between $1.4 \times 10^{-7}$ and
$1.3 \times 10^{-4}$\,\Msun\,yr$^{-1}$ (see
Table~\ref{masslossrates}).  The quoted gas mass-loss rates are
however often based on one diagnostics (e.g., the integrated
intensity of one rotational transition of CO) and only trace a
spatially limited region. To clarify our understanding on the
mass-loss process, it is crucial to study different tracers of mass
loss covering a geometrically extended part of the CSE. We therefore
have performed an in-depth analysis of \emph{ the CO $J$\,=\,1$-$0
through 7$-$6 rotational line profiles in combination with a modeling
of the spectral energy distribution (SED)}. We demonstrate that this
combined analysis gives us the possibility to pinpoint the (variable)
mass loss of \object{WX Psc}.

In Sect.~\ref{observations} we describe the data sets used in this
paper. The method of analysis is outlined in Sect.~\ref{analysis}, and
the results from both the CO line profile modeling and SED modeling
are presented in Sect.~\ref{results}. The results are discussed in
Sect.~\ref{discussion} and are compared with results obtained in other
studies. The conclusions are formulated in Sect.~\ref{conclusions}.

\section{Observations}
\label{observations}
\begin{figure*}
  \sidecaption
  \includegraphics[width=12cm]{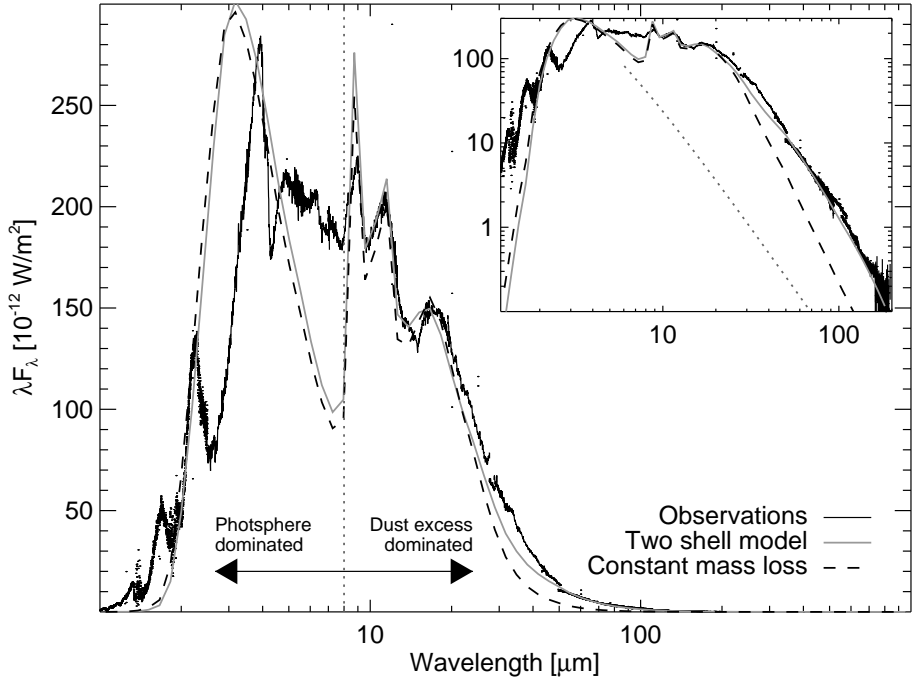}
  \caption{SED of \object{WX Psc} (black), together with the best fit
    dust model (gray) and the best fitting constant mass-loss rate
    model (dashed). The inset shows the same SEDs using logarithmic
    axes in order to emphasize the far-IR slope. Note that the large
    discrepancies between the observed and modeled SEDs between
    1$-$8\,$\mu$m are due to the representation of the stellar
    atmosphere by a blackbody.  The wavelength range which is
    dominated by (unmodeled) photospheric features is to the left of
    the dotted line at 8\,$\mu$m. The main absorption and emission
    features in the 1$-$8\,$\mu$m range are due to H$_2$O, SiO and
    CO. The dust absorption and emission is best traced beyond
    8\,$\mu$m (see text for more details). The dotted line in the
    inset gives an indication of the level of the photospheric
    contribution to the total flux.}
  \label{fig:sed}
\end{figure*}

\begin{table*}
  \begin{center}
    \caption{Mass-loss rates for \object{WX Psc} as found in
      literature. Data used in this paper are indicated with the
      superscript $^{\dagger}$; the appropriate telescope diameter (in
      meter) and beam size (in $''$) are listed in the last column.}
      {\footnotesize{ \setlength{\tabcolsep}{0.5mm}
        \begin{tabular}{lllll}\hline \hline
\rule[-3mm]{0mm}{8mm}Observ. data                                 & \Mdot\ [\Msun\,yr$^{-1}$]             & 
          Comments                                     & Ref.                                  & Diam. -- beam                                                                                                  \\ 
          \hline 
\rule[0mm]{0mm}{5mm}CO (1$-$0)                                     & $1.2 \times 10^{-5}$                  & from analytical expression                       & 
          \citet{Knapp1985ApJ...292..640K}             & 7m -- $100''$                                                                                                                                          \\
          CO (1$-$0)                                     & $2.4 \times 10^{-5}$                  & theor. model of
          \citet{Morris1980ApJ...236..823M}            & 
          \citet{Sopka1989AA...210...78S}              & 20m -- $42''$                                                                                                                                          \\
          CO (1$-$0)                                     & --                                    &                                                  & \citet{Margulis1990ApJ...361..673M}         & 14m -- $45''$ \\
          CO (1$-$0)                                     & --                                    &                                                  & \citet{Nyman1992AAS...93..121N}$^{\dagger}$ & 20m -- $42''$ \\
          CO (1$-$0)                                     & $8.5 (\pm 2.39) \times 10^{-6}$       & from analytical expression                       & \citet{Loup1993AAS...99..291L}              &               \\
          CO (1$-$0)                                     & $1.0 \times 10^{-5}$                  & self-consistent $T(r)$ +
          NLTE radiat. transfer                        & \citet{Justtanont1994ApJ...435..852J} &                                                                                                                \\
          CO (2$-$1)                                     & $2.6 \times 10^{-5}$                  & from analytical expression                       & 
          \citet{Knapp1982ApJ...252..616K}             & 10m -- $30''$                                                                                                                                          \\
          CO (2$-$1)                                     & $1.0 \times 10^{-5}$                  & large difference between data and
          model                                        & \citet{Justtanont1994ApJ...435..852J} &                                                                                                                \\
          CO (2$-$1)                                     & $5.0 \times 10^{-5}$                  & from analytical expression                       & 
          \citet{Jura1983ApJ...275..683J}              &                                                                                                                                                        \\
          CO (2$-$1),(3$-$2),                              & $0.14 - 8.0 \times 10^{-6}$           & from model \citet{Justtanont1994ApJ...435..852J} & 
          \citet{Kemper2003AA...407..609K}$^{\dagger}$ & 15m -- (see text)                                                                                                                                      \\
          (4$-$3),(6$-$5),(7$-$6)                            &                                       &                                                  &                                             &               \\
          L - [12$\mu$m]                               & $2.0 \times 10^{-5}$                  & dust modeling                                   & 
          \citet{Kemper2002AA...384..585K}             &                                                                                                                                                        \\
          OH maser                                     & $2 \times 10^{-5}$                    & using Eq.~(3) in
          \citet{Bowers1983ApJ...274..733B}            & \citet{Bowers1983ApJ...274..733B}
                                                       &                                                                                                                                                        \\
          $J$, $H$, $K^{\prime}$ +                     & $1.3 - 2.1 \times 10^{-5}$            & spherical dusty
          wind with                                    & 
          \citet{Hofmann2001AA...379..529H}            &                                                                                                                                                        \\
          SED + ISI (11\,$\mu$m)                       &                                       & $\rho \propto r^{-1.5}$ from 137\,\Rstar\
          on                                           &                                       &                                                                                                                \\
          $J$, $H$, $K^{\prime}$ +                     & $1.3 \times 10^{-4}$                  & spherical dusty wind
                                                       & \citet{Tevousjan2004ApJ...611..466T}                                                                                                                   \\  
            ISI (11\,$\mu$m)                           &                                       &                                                  &                                                             \\
          $J$, $H$, $K^{\prime}$ +                     & $9 \times 10^{-6}$  last 550\,yr      & spherical
          dusty wind + cocoon                          & \citet{Vinkovic2004MNRAS.352..852V}   &                                                                                                                \\
\rule[-3mm]{0mm}{3mm}SED                                          & $2 \times 10^{-5}$,  550 -- 5500 \,yr & 
                                                       &                                       &                                                                                                                \\
          \hline
        \end{tabular}
      }}
    \label{masslossrates} 
  \end{center}
\end{table*}

The analysis is based on \emph{(i)} rotational CO line profiles found
in the literature and \emph{(ii)} the SED in the range from 0.7 to
1300\,$\mu$m. The applied CO-dataset is indicated with the superscript
$\dagger$ in Table~\ref{masslossrates}. Main emphasis is put on the
high-quality data obtained by \citet{Kemper2003AA...407..609K} on the
15\,m \emph{JCMT} telescope on Mauna Kea, Hawai. The half power beam
width of the different receivers was $19.7''$ for CO (2$-$1), $13.2''$
for CO (3$-$2), $10.8''$ for CO (4$-$3), $8.0''$ for CO (6$-$5), and
$6.0''$ for CO (7$-$6).  The absolute flux calibration accuracy was
estimated to be 10\,\% for the CO (2$-$1), (3$-$2) and (4$-$3), and
${\sim}$30\,\% for the CO (6$-$5) and (7$-$6) line because the flux
standards, that are available for these higher excitation lines, are
less reliable. We note that in principal pointing uncertainties,
beam effects, atmospheric seeing, and systematics at these frequencies
can conspire against you to produce an error that is larger than
30\,\%. Unfortunately we don't have enough statistics information on
standards etc.\ to know for sure how bad a problem this was for the
RxE and RxW/D receivers at the time of the observations.

The SED of \object{WX Psc} is that of a strongly reddened Mira-type
variable star with a period of 660 days and a K band amplitude of 3.4
mag \citep{Samus2004}. Two full spectral scans from both spectrometers
(Short Wavelength Spectrometer  (\emph{SWS}) and Long Wavelength
Spectrometer (\emph{LWS})) onboard the Infrared Space Observatory
(\emph{ISO}) satellite are available. Due to the pulsation of the star
the flux levels of the SWS spectra differ by a factor $\sim$2. We opt
to use the SWS spectrum taken on 15 Dec 1996 (tdt 39502217) and the
LWS spectrum from 15 Jun 1997 (tdt 57701103).  Although these spectra
have not been obtained simultaneously, the flux levels in the
overlapping region between SWS and LWS agree reasonably well because
the observing dates correspond to phases 0.38 and 0.66, i.e. roughly
symmetric compared to the minimum at 0.5. We have further used the
0.97$-$2.5 $\mu$m near-IR spectrum from
\citet{Lancon2000A&AS..146..217L} scaled to the flux-level of the SWS
spectrum. The IR spectroscopic data have been supplemented by optical
and IR photometric data \citep{2MASS, Dyck1974ApJ...189...89D,
Beichman1988iras....1.....B, Smith2004ApJS..154..673S,
Morrison1973ApJ...186..193M, Hyland1972A&A....16..204H}, sub-mm
photometry \citep{Sopka1989AA...210...78S} and radio measurements
\citep{Walmsley1991A&A...248..555W, dehaes} in order to construct the
full SED (see Fig.~\ref{fig:sed}).

In the wavelength region up to $\sim$8\,$\mu$m, photospheric
 absorption features and emission features arising from density
 enhancements in the outer atmosphere caused by pulsational levitation
 are visible. Specifically, the CO first overtone and H$_2$O $\nu_1$
 symmetric stretch and $\nu_3$ asymmetric stretch band features are
 arising between $\sim 2$ and $3.8$\,$\mu$m, the small absorption peak
 around 4.05\,$\mu$m\ is caused by the SiO first overtone, with the
 stronger absorption peak at slightly larger wavelengths (at 
 $\sim$5\,$\mu$m) mainly being due to the CO fundamental
 vibration-rotation band. From 6.27\,$\mu$m onwards, the H$_2$O
 $\nu_2$ vibrational bending band is the main absorption feature.
 However, for the purpose of determining the mass-loss history of
 \object{WX Psc}, a detailed analysis of the atmospheric behaviour of
 this target is not a prime issue. In the framework of the dust
 modeling presented here, we simply represent the central star by a
 blackbody. The assumption of a blackbody for the underlying
 star does not influence the conclusions since dust emission and
 absorption dominates over photospheric effects beyond 8\,$\mu$m (see
 Fig.~\ref{fig:sed}). We focus on several \emph{dust} characteristics
 of the SED and IR spectroscopy to estimate the mass loss. The most
 significant of these are {\it 1)} the average reddening of the star;
 {\it 2)} the self-absorption of the 10\,$\mu$m silicate feature; {\it
 3)} the relative strength of the 10 and 20\,$\mu$m bands, and {\it 4)}
 the slope in the mid- to far-IR. These observables trace the relative
 amounts of dust near and further away from the star, i.e.\  the
 columns of warm and cooler material along the line of sight.

\section{Analysis}
\label{analysis}
To constrain the dust-driven wind structure from its base 
out to several thousands of stellar radii, we use both 
molecular line fitting of the CO rotational $J = 1-0$ through 7$-$6 line
profiles and  dust radiative transfer simulations of the
global SED and IR spectrum. The spectral lines are modeled using
GASTRoNOoM, the dust continuum using {\sc modust}. We briefly discuss both
programs in Sect.~\ref{codes}. Special emphasis is
given on the model selection procedure and on the measure of the
goodness-of-fit in Sect.~\ref{statistics}. The results are presented
in Sect.~\ref{results}.

\subsection{Theoretical code}
\label{codes}
\subsubsection{GASTRoNOoM} 
\label{gastronoom} 
The observed line profiles provide information on the thermodynamical
structure of the outflow of \object{WX Psc}. For a proper
interpretation of the full line profiles, we have developed a
theoretical model (GASTRoNOoM - Gas Theoretical Research on Non-LTE
Opacities of Molecules) which first (1) calculates the kinetic
temperature in the shell by solving the equations of motion of gas and
dust and the energy balance simultaneously, then (2) solves the
radiative transfer equation in the co-moving frame
\citep[CMF,][]{Mihalas1975ApJ...202..465M} using the Approximate
Newton-Raphson operator as developed by
\citet{Schonberg1986A&A...163..151S} and computes the NLTE
level-populations consistently, and finally (3) determines the
observable line profile by ray-tracing. A full description of the code
is presented in \citet{DecinCOI}.  The main assumption is a
spherically symmetric mass loss. The mass-loss rate is allowed to vary
with radial distance from the star. Note that the gas and drift
velocity are not changed accordingly. Adopting a value for the gas
mass-loss rate, the dust-to-gas ratio is adjusted until the observed
terminal velocity is obtained.  For details on the thermal balance,
the gas and dust velocity, and the treatment of variable mass loss we
refer to \citet{DecinCOI}. We already note that for solving the
radiative transfer equation in the CMF-frame in the case of \object{WX
Psc}, we have used a value for the mass-loss rate, \Mdot, equal to $1
\times 10^{-5}$\,\Msun/yr.

\subsubsection{MODUST} 
\label{modust} 
We have used the code {\sc modust} \citep{bouwman00} to model the
transfer of radiation through the dusty outflow of \object{WX Psc}.
This code has been applied extensively to model not only the dust in
the outflows of supergiants and AGB stars
\cite[e.g.][]{voors00,kemper01,dijkstra03,hony03,dijkstra06}, but also
solid state particles in proto-planetary disks and comets
\cite[e.g.][]{bouwman03}. In short, the code assumes a spherical
distribution, such as an outflow or shell, of dust grains in radiative
equilibrium that are irradiated by a central star. Here, we represent
the photospheric energy distribution of \object{WX Psc} by a blackbody
(see Sect.~\ref{observations}). A special circumstance in the modeling
of \object{WX Psc} is that we account for two distinct dust shells
(see Sect.~\ref{SEDfitting}). For the inner shell we do a consistent
radiation transfer solution, for the outer shell we assume that the
dust medium is optically thin and that it is irradiated by the light
emerging from the inner shell. We have verified that indeed the outer
shell of \object{WX Psc} is optically thin.

\subsubsection{Parameter estimation, model selection, and the assessment of goodness-of-fit}
\label{statistics} 
The observational line data are subject to two kinds of uncertainties:
(1) the \emph{systematic} or \emph{absolute} errors, with variance
$\sigma^2_{{\rm abs}}$, arising e.g.\ from uncertainties in the
correction for variations in the atmospheric conditions, and (2) the
\emph{statistical} or \emph{random} errors $\varepsilon$, with
variance $\sigma^2_{{\rm stat}}$, which reflect the variability of the
points within a certain bin. Both kind of errors have to be taken into
account when estimating the model parameters. In case of the CO line
profiles, the statistical errors (rms) are smaller than the systematic
errors. We hence will construct a statistical method giving a greater
weight to the line profile than to the integrated line
intensity. Moreover, as demonstrated in \citet{DecinCOI}, especially
the line profiles yield strong diagnostics for the determination of
the mass-loss \emph{history}.

To determine the mass-loss history of \object{WX Psc} from the CO line
profiles, a grid of ${\sim}$300\,000 models was constructed. In a
first step, the models for which the predicted line profiles did not
fulfill the absolute-flux calibration uncertainty criterion were
excluded. This absolute flux criterion does not only include the
absolute flux uncertainty on the data as specified in
Sect.~\ref{observations}, but also incorporates the noise on the
data. {F}rom then on, the other models were treated equally when
judging the quality of the line profile \emph{shape}. To do so, a
scaling factor based on the ratio of the integrated intensity of the
observed data to the integrated intensity of the predicted data was
introduced to re-scale the observational data. {F}rom a statistical
standpoint, this scaling factor merely is an additional model
parameter, accounting for the calibration uncertainties.

Assume a model for the re-scaled observed spectrum $y_i$ at velocity
$i$, $i=1,2,\dots,n$ with expected mean value ${\rm{E}}(y_i) = \mu_i$,
representing the theoretical spectrum of the target. The statistical
measurement errors are assumed to be normally distributed with mean 0
and variance taken to be constant at all frequency points $i$, i.e.\
$\sigma_{{\rm stat}_i} \equiv \sigma_{\rm stat}$. Hence, the model is
assumed to follow a normal distribution
\begin{equation}
\label{like0}
y_{i}=\mu_{i}+\varepsilon_{i},\ \  {\rm where\ } \varepsilon_{i}\sim
N(0,\sigma^{2}_{\rm stat}). 
\end{equation}
A key tool in model selection and fitting, and in the assessment of
goodness-of-fit, is the log-likelihood function which, for the
normally distributed case, takes the form $\ell$, defined
as:
\begin{eqnarray}
  \ell = \sum_{j=1}^{N_{\rm lines}} \left( \sum_{i=1}^{n_j}
    \left[ -\ln(\sigma_{{\rm stat}_j}) - \ln(\sqrt{2\pi}) -
      \frac{1}{2} \left( \frac{y_{j,i} - \mu_{j,i}}{\sigma_{{\rm
              stat}_j}} \right)^2 \right] \right),
\end{eqnarray}
where $N_{\rm lines}$ represents the total number of rotational lines
being 6, and $n_j$ the number of frequency points in the $j$-th line
profile \citep{Cox1990}. The model which \emph{best} fits the observed data
\emph{maximizes} the log-likelihood distribution.

\begin{table}
  \caption{$(1-\alpha)\,100$\,\% quantile of the 
    $\chi^2_p$-distribution, with $p$ degrees of freedom and a
    significance level $\alpha = 0.05$.}
  \begin{center}
    \begin{tabular}{rl|ll} \hline
      \rule[-3mm]{0mm}{8mm} $p$ & $\chi^2_p(\alpha)$ & $p$ &
      $\chi^2_p(\alpha)$ \\ 
      \hline
      1                         & 3.8414588                & 11  & 19.675138 \\
      2                         & 5.9914645                & 12  & 21.026070 \\
      3                         & 7.8147279                & 13  & 22.362032 \\
      4                         & 9.4877290                & 14  & 23.684791 \\
      5                         & 11.070498                & 15  & 24.995790 \\
      6                         & 12.591587                & 16  & 26.296228 \\
      7                         & 14.067140                & 17  & 27.587112 \\
      8                         & 15.507313                & 18  & 28.869299 \\
      9                         & 16.918978                & 19  & 30.143527 \\
      10                        & 18.307038                & 20  & 31.410433 \\
      \hline
    \end{tabular}
  \end{center}
  \label{critical_numbers_chi2} 
\end{table}

One can derive three properties from the log-likehood distribution
\citep[see][]{Welsh1996}.  First, for a given parametric model,
maximizing $\ell$ leads to the so-called maximum likelihood estimator
$\widehat{\bftheta}$ for the model parameters. Denote by $\ell_m$ the
value of the log-likelihood function at maximum.  Standard errors are
deduced as the square root of the diagonal of the negative inverse
information matrix. The information matrix is the matrix of second
derivatives of the log-likelihood function w.r.t.\ the model
parameters.

Second, the log-likelihood function can be used to determine
$(1-\alpha)\,100\,\%$ confidence intervals for the model parameters,
where $\alpha$ is the significance level. Precisely this is done
through the method of profile likelihood, and the corresponding
intervals are termed `profile likelihood confidence intervals'. Let us
focus on one of the parameters, $\varphi$ say. The profile likelihood
interval is defined as the values $\varphi$ for which
$\ell_{\varphi}\equiv\ell(\varphi,\widehat{\bftheta}(\varphi))$ is
sufficiently close to $\ell_m$. We use $\widehat{\bftheta}(\varphi)$
as the parameter that maximizes the log-likelihood function subject to
the constraint that $\varphi$ is kept fixed.  Sufficiently close is to
be interpreted in the precise sense that $2(\ell_m-\ell_{\varphi})\le
\chi^2_1(\alpha)$, where $\chi^2_1(\alpha)$ is the $(1-\alpha)\,100\%$
quantile of the $\chi^2_1$ distribution.  The same procedure can be
used for a group of $p$ parameters simultaneously. In this case
$\varphi$ is $p$-dimensional and the quantile is now of the
$\chi^2_p(\alpha)$-distribution (see
Table~\ref{critical_numbers_chi2}).  In our case, the confidence
intervals are influenced by the grid spacing, but they converge to the
true intervals if the grid spacing becomes arbitrarily fine.

Third, the log-likelihood function can be used to compare two
different models, with a different number of parameters, $p$ and
$p+r$, as long as they are nested. The more elaborate model is
preferred if $2(\ell^{p+r}_m-\ell^{p}_m)\ge\chi^2_r(\alpha)$,
otherwise preference points towards the simpler model.

 One should realize that the outlined log-likelihood method does
  not force the accordance between observations and predictions to be
  better than a certain pre-specified level in order to be
  acceptable. This is in contrast with the often used
  reduced-$\chi^2$-method, which requires that for a good fit the
  value of the reduced-$\chi^2$ should be lower than 1 (or 2). Using
  the log-likelihood method, one is selecting the theoretical
  predictions (and corresponding confidence intervals for the model
  parameters) with the best fit to the data in the framework
  of the applied theoretical model.

As an example, we describe the selection of the best model and the
determination of the 95\,\% confidence intervals of the stellar and
mass-loss parameters as derived from the CO line fitting procedure
(see Sect.~\ref{sec:modell-co-rotat}).  In Fig.~\ref{WXPsc_mdot}
an example is shown of a mass-loss profile with 4 \Mdot-phases and 3
small interfaces (each described by 2 parameters ($R_i$,
\Mdot$_i$)). This mass-loss profile is hence described by 6 pairs
($R_i$, \Mdot$_i$), indicated with diamonds in
Fig.~\ref{WXPsc_mdot}. Other free parameters in the grid computation
were the stellar temperature $T_{\star}$, dust condensation radius
$R_{\rm{inner}}$, and outer radius of the CSE $R_{\rm{outer}}$,
resulting in 15 free parameters in total.  We note that for the
modelling of \object{WX Psc} the value of \Mdot\ at $R_{\rm{inner}}$
was specified to stay constant till the first decrease in mass loss at
$\sim$30\,\Rstar, and also from 1550\,\Rstar\ on the mass-loss was
kept constant until $R_{\rm{outer}}$. This is however not a
prerequisite to the code, but has been introduced to reduce the grid
computations. Our best-fitting model has a log-likelihood $\ell_m =
-314.17$. The 95\,\% confidence intervals are determined by all
models whose log-likelihood $\ell$ is $2(\ell_m - \ell)\le
\chi^2_{15}(\alpha)$, i.e.\ $2(\ell_m - \ell)\le 24.99579$ or
$-326.6 \le \ell \le -314.17$. If we want to test if a model with 3
episodes of high mass loss yields a significantly better result
than with 2 high-mass epochs, we are introducing 4 extra pairs
($R_i$, \Mdot$_i$), i.e.\ 8 free parameters. Hence, the more elaborate
model with 23 parameters is preferred when the maximum of the
log-likelihood of the 3-shell model ($\ell_m^{23}$) obeys
$2(\ell^{23}_m-\ell^{15}_m) \ge \chi^2_8(\alpha)$ or $\ell_m^{23} \ge
-306.41$.

\section{Results}
\label{results}
\subsection{Modeling the CO rotational line profiles}
\label{sec:modell-co-rotat}
\begin{figure}
  \includegraphics[height=8.8cm,angle=90]{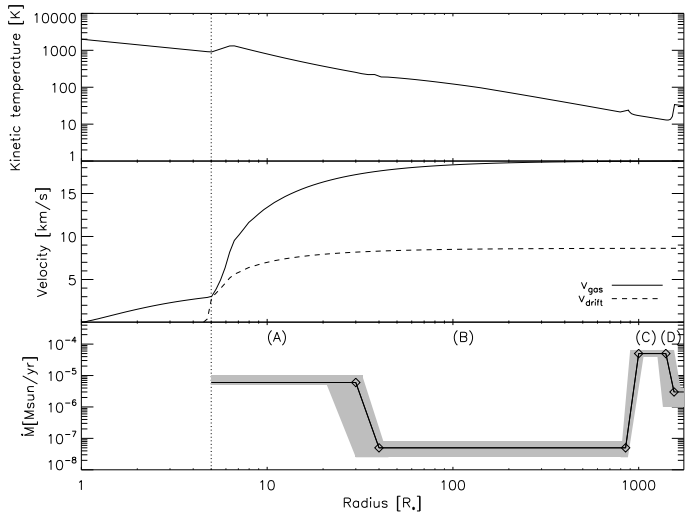}
\vspace*{0.5ex}
  \caption{\emph{Upper:} Estimated temperature profile, \emph{middle:}
    estimated velocity structure, and \emph{bottom:} estimated
    mass-loss rate \Mdot$(r)$ for \object{WX Psc} as a function of
    radial distance to the star (black line). The shading region in
    the bottom panel displays the confidence intervals of the
    estimated mass-loss parameters.}
  \label{WXPsc_mdot}
\end{figure}

\begin{table}
  \setlength{\tabcolsep}{0.05mm}
  \caption{Parameters of the model with best goodness-of-fit for
    \object{WX Psc}. In the second column models parameters as
    derived from the CO rotational lines are listed, in the third
    column the parameters as estimated from the SED fitting are
    given. The numbers in italics are input parameters that have been
    kept fixed at the given values. There is, of course, an
    interdependence between $T_{\star}$, $R_{\star}$ and $L_{\star}$
    and between $L_\star$, the distance, and $m_{bol}$ from which
    the distance has been derived.} 
  \begin{tabular}{lll}\hline \hline
    parameter                       & from CO lines                                    & from SED fitting                                \\
    \hline
    $T_{\star}$ [K]                 & 2000                                                   & \emph{2000}                                            \\
    $R_{\star}$ [$10^{13}$\,cm]     & 5.5                                                    & \emph{5.5}                                             \\
    $L_{\star}$ [$10^4$\,\Lsun ]    & \emph{1}                                                & \emph{1}                                               \\
    dust-to-gas ratio               & 0.004                                            & \emph{0.004}                                           \\
    $\varepsilon$(C) [$10^{-4}$]    & \emph{3}                                                & $-$                                             \\
    $\varepsilon$(O) [$10^{-4}$]    & \emph{8.5}                                              & $-$                                             \\
    distance         [pc]           & \emph{833}
    &       \emph{833}                                             \\
    $R_{\rm{inner}}$ [\Rstar]       & 5                                                & 5                                               \\
    $R_{\rm{outer}}$ [\Rstar]       & 1750                                             & 1500$^{\dagger}$                                \\
    $v_{\infty}$     [km\,s$^{-1}$] & \emph{18}                                               & \emph{18}                                              \\
    \Mdot$(r)$ [{\Msun}yr$^{-1}$]\ \ \ (A)   & ${\sim}6 \, 10^{-6}$ ($\sim$5$-$30\,\Rstar)      & ${\sim}4 \, 10^{-5}$ ($\sim$5$-$50\,\Rstar)   \\ 
\phantom{\Mdot$(r)$ [{\Msun}yr$^{-1}$]}\ \ \ (B)  & ${\sim}5 \, 10^{-8}$ ($\sim$30$-$850\,\Rstar)    & $-$                                             \\ 
\phantom{\Mdot$(r)$ [{\Msun}yr$^{-1}$]}\ \ \   (C)  & ${\sim}5 \, 10^{-5}$ ($\sim$850$-$1450\,\Rstar)  & ${\sim}3 \, 10^{-4}$ ($\sim$800$-$1500\,\Rstar) \\ 
\phantom{\Mdot$(r)$ [{\Msun}yr$^{-1}$]}\ \ \   (D)  & ${\sim}3 \, 10^{-6}$ ($\sim$1450$-$1750\,\Rstar) & $-$                                             \\ 
    \hline
  \end{tabular}
  \label{param_WXPsc} 
{\footnotesize{$^{\dagger}$ The outer radius mentioned in the SED fitting refers to
  the outer edge of the second shell, called region (C) in
  Fig.~\ref{WXPsc_mdot}, and should be compared to the 1450\,\Rstar\
  value derived from the CO modeling.}}
\end{table}

\begin{figure*}
  \sidecaption
  \includegraphics[height=\textwidth,angle=90]{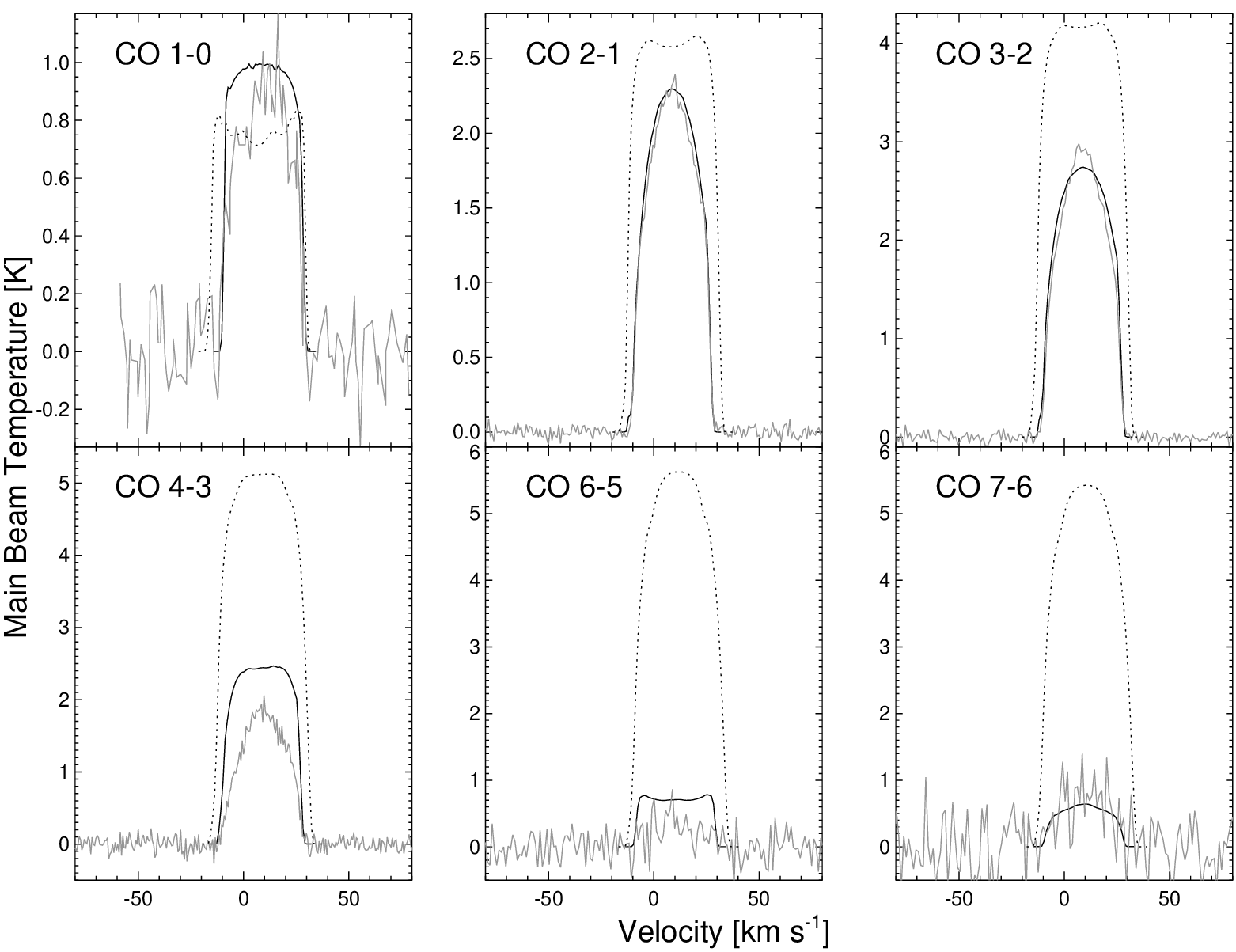}
\vspace*{2ex}
  \caption{CO rotational line profiles of \object{WX Psc} \citep[full
    grey line;][]{Nyman1992AAS...93..121N, Kemper2003AA...407..609K}
    compared with \emph{(i)} black dotted line: the model predictions based
    on the parameters as given by
    \citet{Justtanont1994ApJ...435..852J}, who assumed a constant mass
    loss of \Mdot\,=\,$1 \times 10^{-5}$\,\Msun\,yr$^{-1}$ and a
    distance of 850\,pc, and \emph{(ii)} full black line: the best-fit model
    with parameters as specified in Table~\ref{param_WXPsc}. 
  \label{WXPsc_model} }
\end{figure*}

Comparing the observed CO line profiles with theoretical profiles
predicted by a model assuming a constant mass loss confirms the
hypothesis proposed by \citet{Kemper2003AA...407..609K} that variable
mass loss may play an important role in the shaping of the spectral
line profiles (see dotted line in Fig.~\ref{WXPsc_model}).

Using the procedure as outlined in Sect~\ref{statistics}, a total
amount of ${\sim}$300\,000 models was run to estimate the mass-loss
history of \object{WX Psc}. A luminosity of $1 \times 10^4$\,\Lsun\
was assumed, which together with the integrated luminosity of the SED
yields a distance of 833\,pc. The expansion velocity traced by all
line profiles is $v_{\infty} = 18 \pm 1$\,\kms. For the carbon and
oxygen abundances in the outer atmosphere of the target, we opt to use
the cosmic abundance values of \citet[][see
Table~\ref{param_WXPsc}]{Anders1989GeCoA..53..197A}, and it is
assumed that in case of the oxygen-rich target \object{WX Psc} all of
the available carbon will be locked in CO\footnote{Note that in case
of a variable mass loss, the [CO/H$_2$]-ratio is assumed to follow the
results of \citet{Mamon1988ApJ...328..797M}. Since the total
abundance of the main molecule H$_2$ varies according to the
\Mdot-profile, the total CO abundance varies accordingly.}.  Other
model parameters as stellar temperature \Tstar\ (or stellar radius
\Rstar), dust condensation radius $R_{\rm inner}$, outer radius of the
envelope $R_{\rm outer}$ and the mass-loss profile were kept free in
the grid running.  In case of constant mass loss, the value of $R_{\rm
outer}$ is set at the radius where the CO abundance drops to 1\,\% of
its value at the photosphere, using the photodissociation results of
\citet{Mamon1988ApJ...328..797M} \citep[see][]{DecinCOI}. However, the
CO photodissociation radius by \citet{Mamon1988ApJ...328..797M} were
obtained in case of constant mass loss, and it is questionable how
this applies when dealing with variable mass loss. In case of
variations in the mass loss, we therefore opt to have $R_{\rm outer}$
as free parameter, with the additional constraint that $R_{\rm outer}$
should be smaller than the $R_{\rm outer}$-value obtained using the
formula of \citet{Mamon1988ApJ...328..797M} for a value of the mass
loss equal to the mass-loss rate at the outermost radius point. We
hereby want to ensure not to go beyong the photodissociation radius as
calculated by \citet{Mamon1988ApJ...328..797M}. Moreover, it is always
checked that the full line forming region \citep[as traced by $I(p)
  \times p^3$, see e.g.\ Fig.~5 and 6 in][]{DecinCOI} of all
rotational CO lines lies within $R_{\rm{outer}}$.

Parameters characterizing the CSE of the model with the best
goodness-of-fit are specified in the second column of
Table~\ref{param_WXPsc}. The derived temperature profile, velocity
structure and mass-loss history are displayed in
Fig.~\ref{WXPsc_mdot}; a comparison between observed and theoretical
line profiles is shown in Fig.~\ref{WXPsc_model}.  We note that the
abruptness of changes in the mass-loss profile causes the small-scale
ripples seen in the CO line profiles\footnote{Theoretical profiles are
  calculated with 150 impact paramters and 200 quadrature point in the
$I_{\nu}\,p\,dp$-integral.}. The presence of the
noise in the line profile does however not influence the conclusions
in this paper. The dust-to-gas
ratio is derived from the mass-loss rate and terminal velocity and
equals 0.004. This value for the dust-to-gas ratio is quite uncertain.
As described in \citet{DecinCOI}, this value is taken constant
throughout the entire envelope. However, a higher mass-loss rate as in
region (C) can drive a dusty wind with a different velocity than in a
low-mass region and perhaps also with a different dust-to-gas ratio.
In this respect, we note that the different CO lines, originating from
various parts in the wind, do not exhibit different line widths
(Fig.~\ref{WXPsc_model}).

Before comparing with the results of the SED fitting
(Sect.~\ref{SEDfitting}), we first discuss the derived \Mdot-profile
and associated uncertainties (see bottom panel in
Fig.~\ref{WXPsc_mdot}). Acceptable values for the mass-loss rate in
region (A) range from $5 \times 10^{-6}$ to $1 \times
10^{-5}$\,\Msun\,yr$^{-1}$, for region (B) from $2.5\times 10^{-8}$ to
$8 \times 10^{-8}$\,\Msun\,yr$^{-1}$, for region (C) from $4 \times
10^{-5}$ to $6.5 \times 10^{-5}$\,\Msun\,yr$^{-1}$, and for region (D)
from $1 \times 10^{-6}$ to $4 \times 10^{-6}$\,\Msun\,yr$^{-1}$.
Especially region (C) with enhanced mass loss is constrained quite
well since the (high-quality) low-excitation lines are mainly formed
in this region. Notice that the temperature increase in region D (see
top panel in Fig.~\ref{WXPsc_mdot}) is quite uncertain: an altered
temperature structure with a continuously decreasing temperature
reaching 10\,K at the outer radius, yields line intensity profiles
which are better for all low-excitation transitions, and has a
log-likelihood that is significantly better than that of our
`best-fit' model with an amount $\Delta \ell = 105$.  The main cause
of uncertainty in $T(r)$ is that interactions between the different
shells and with the ISM are not taken into account.

Inspecting Fig.~\ref{WXPsc_model}, we see a good resemblance between
the observations and model predictions for the CO (1--0), (2--1) and
(3--2) lines. The predicted peak intensity of the CO (4--3) and (6--5)
line is somewhat too high. This may indicate that the temperature in
region C, where the dominant part of these lines is formed, may be
somewhat lower than is inferred.

Almost half of the models was run with a third shell extending between
$\sim$100 and $\sim$600\,\Rstar, and with mass-loss rates ranging
between $2 \times 10^{-8}$ and $3 \times
10^{-5}$\,\Msun\,yr$^{-1}$. The model with the best fit had
\Mdot\,=\,$5 \times 10^{-8}$\,\Msun\,yr$^{-1}$ both in region (B) and
in this extra third shell, i.e. the statistical analysis gives
preference to a 2-shell model above a 3-shell model. Models still
having an acceptable fit to the line profiles, all had \Mdot\,$\le 2.5
\times 10^{-7}$\,\Msun\,yr$^{-1}$ in this third shell.

\subsection{SED fitting} 
\label{SEDfitting}
The SED of the best fitting model is presented in Fig.~\ref{fig:sed}
and the implied model parameters are given in Table~\ref{param_WXPsc}.
As described above, the model consists of two shells that present two
episodes of high mass loss. The intermediate period of significantly
lower mass loss is not modeled. In general, we find that the wind
structure that best reproduces the observed SED has very similar
parameters to those derived from modeling the CO rotational lines. In
particular, the two methods agree on the location of the inner edges
of the two shells. The outer radii of the shells can be chosen to be
roughly consistent with the values derived from the molecular
modeling. It is impossible to derive statistical uncertainties on
the derived parameters since the best fitted model is not obtained
through a formal fitting procedure (like was done for the CO lines). Such
a formal fitting procedure is inhibited by the very unequal weighting
that is given to the various regions of the SED. For example, the
region which is dominated by water vapour emission and absorption is
virtually neglected. While the integrated photospheric component is
used as a constraint, no single optical wavelength range is given
priority. In contrast the 8 -- 25\,$\mu$m range, which exhibits the
clearest dust spectral signatures, is inspected in detail.
Furthermore, the IR-submm slope is used as a strong constraint
eventhough it is determined by far fewer points than the mid-IR part.
However we can determine by how much we can vary the parameters
without significantly reducing the quality if the fit. Varying, e.g.,
the inner radii of the two shell model by $\sim$10 per cent,
significantly degrades the quality of the fit. We also note that the
dust model results are less sensitive to the precise extent of the
dust shell than to the amount of mass contained in the
shell. Therefore, the outer radii are less well constrained. An
important conclusion that can be drawn is that the self-absorption and
relative strengths of the 10 and 20\,$\mu$m silicate features arise
from a dense and compact shell close to the photosphere. To illustrate
this point we also show in Fig.~\ref{fig:sed} the ``best'' fitting
model using a single shell from a constant mass loss ($\dot{M}$=
1.5\,10$^{-5}$\,M${_\odot}$; $R$=5$-$200 R$_\odot$). A more extended
shell would yield too much cold dust and a less dense wind would not
yield enough self-absorption. This compact shell is however not
sufficient to explain the far-IR data. The slope of the SED in the
far-IR is dominated by cool material (see inset in
Fig.~\ref{fig:sed}).

\subsection{CO versus SED fitting}
\label{comparisonCOSED}
 The fact that these very different tracers yield such similar
  parameters with respect to the \emph{location} of the densest shells
  is remarkable and builds further confidence in the reality of the
  mass-loss variabilities. However, it should be noted that the
  mass-loss \emph{rates} derived from fitting the SED (column~3 in
  Table~\ref{param_WXPsc}) are higher than those found from modeling
  the CO emission (column~2 in Table~\ref{param_WXPsc}), by roughly a
  factor 6. Partly, the discrepancy can be accounted for by
  considering the material contained in the regions that are not
  included in the dust model (B and D). Some experimenting with the
  dust modeling shows that the differences can not be reconciled by
  adapting the dust composition alone. Upon perusal of
  Table~\ref{masslossrates}, it appears that those studies that take
  the thermal IR emission into account systematically derive a denser
  wind than those using gas tracers alone. The difference may be
  related to the assumed CO/H$_2$ ratio (see
  Table~\ref{param_WXPsc}). Under the assumption that all carbon is
  locked in CO, the CO/H$_2$-ratio used in this paper is
  $3\,10^{-4}$. \citet{Zuckerman1986ApJ...304..394Z} used a value of
  $5\,10^{-4}$ for the CO/H$_2$-ratio in O-rich giants, while
  \citet{Knapp1985ApJ...292..640K} reported a value of $3\,10^{-4}$.
  This value differs from AGB star to AGB star, as it is dependent on
  the amount of dredge-up. We estimate the uncertainty on the used
  CO/H$_2$-ratio to be a factor 2. In addition, as described in
  Sect.~\ref{sec:modell-co-rotat}, the derived dust-to-gas ratio as
  calculated from the CO modeling is somewhat uncertain. This value is
  not only taken constant throughout the whole envelope, but moreover
  it is assumed that {\it all} the dust particles participate in
  accelerating the wind to the observed outflow velocity. The
  discrepancy in dust content between the two types of models may
  signify that a sizable fraction of the dust, though present, does
  not contribute much to the acceleration of the wind. If some
  fraction of the dust particles is effectively shielded from
  illumination by the star, they will not feel the radiation force
  from the star and, by consequence, will not contribute to the
  acceleration of the wind. This kind of shielding can typically occur
  in randomly distributed density inhomogeneities (usually referred to
  as clumps) or can be due to deviations of spherical symmetry.  Only
  some fraction of the dust is hence responsible for the driving; via
  collisions with the inert gas, the gas is accelerated, which in
  it turn may accelerate the inert dust. In the acceleration of the
  gas particles, it is not only assumed that all dust particles
  participate in this proces, but also that the dust grains transfer
  all of the momentum they acquire from the radiation field to the gas
  through collisions. Therefore, an alternative explanation could be
  that there is an incomplete momentum coupling between the dust and
  the gas in the wind
  \citep[e.g.][]{MacGregor1992ApJ...397..644M}. Lastly, an explanation
  could have been dust which is not partaking in the outflow but
  present in a stable configuration, i.e.\ a disk.  However, the
  gaseous material present in such a slowly rotating disk would be
  detected via the presence of a very narrow line profile (typically
  line width less than 5\,km\,s$^{-1}$) superimposed on a broad
  outflow component \citep[e.g. as is the case of \object{RV
  Boo},][]{Bergman2000A&A...353..257B}. Since such a narrow component
  is not detected in any of the line profiles this explanation is
  ruled out.

\section{Discussion} 
\label{discussion}
The diagnostic power of estimating the mass-loss history from
modeling the rotational CO line profiles and the full SED, is
evident. Both our CO and SED modeling are however based on one common
assumption, namely a spherically symmetric envelope. We comment on
this assumption in Sect.~\ref{spherical}. In Sect.~\ref{comparison} we
compare our results to other studies. We discuss the time scale of the
mass-loss variability and its implication on the possible mechanism
driving this variability in Sect.~\ref{variability}.

\subsection{Structure of the circumstellar envelope} 
\label{spherical}
The detailed structure of the CSEs of AGB stars and the physics
driving this structure remain unknown after half a century of study.
The main reason being that the cool material is a real challenge to
high-resolution observations. From a survey of AGB envelopes using
millimeter interferometry, \citet{Neri1998A&AS..130....1N} concluded
that most (${\sim}$70\,\%) AGB envelopes are consistent with spherical
symmetry. \object{WX Psc} belongs to the small group of AGB stars
whose CSE has already been studied using different techniques, and for
which several indications are found for deviations from a spherically
symmetric envelope. \emph{(1)} \citet{Neri1998A&AS..130....1N} fitted
the CO (1--0) and (2--1) interferometric observations with a
two-component Gaussian visibility profile consisting of a circular and
an elliptical component. The circular flux-dominant component
corresponds to a spherical envelope with a $29''.6$ diameter, the
elliptical one to an axisymmetric envelope with major axis of $9''.8$
and minor axis of $6''.8$, and a position angle of $-45^{\circ}$ (E to
N). \emph{(2)} Broad-band near-infrared polarimetry performed by
\citet{McCall1980A&AS...42..141M} revealed that \object{WX Psc} is
highly polarized in the $I$ and $J$-band, while the polarization
pattern is not dominant in the $H$ and $K^{\prime}$-band. This may
either indicate an alignment of dust particles or an asymmetrical
dust-shell structure. \emph{(3)} Another technique of imaging the
circumstellar dust is from bispectrum speckle-interferometry. Using
this technique, \citet{Hofmann2001AA...379..529H} showed that the $H$
and $K^{\prime}$-band images of \object{WX Psc} appear almost
spherically symmetric, while the $J$-band shows a clear asymmetry. Two
structures can be identified in the $J$-band image: a compact
elliptical core and a fainter fan-like structure, along a symmetry
axis of position angle $-28^{\circ}$, out to distances of
${\sim}$200\,mas. \emph{(4)} \citet{Mauron2006A&A...452..257M} imaged
the circumstellar dust in scattered light at optical
wavelengths. While the images of the extended envelope appear
approximately circularly symmetric in the $V$-band, the Hubble
Space Telescope (\emph{HST}) images revealed that the inner region of the CSE
consists of a faint circular arc at ${\sim}$0.8\arcsec to the NW and a
highly asymmetric core structure, with a bright extension out to
${\sim}$0.4\arcsec (or 90\,\Rstar\ at 833\,pc) at position angle
$-45^{\circ}$. Most likely, the HST image captures the extension of
the core asymmetry seen in the $J$-band. They concluded that
\object{WX Psc} is somewhat similar to \object{IRC\,+10216} where
\emph{approximate circular symmetry and shells in the extended CSE
co-exist with a strong axial symmetry close to the star}.

Nevertheless, we assume a spherically symmetric geometry for the whole
wind structure when modeling the SED and CO line profiles. The
different \emph{direct} pieces of evidence for deviation of spherical
symmetry all trace small scales ($\la 200$\,mas). On larger scales,
the shell does not appear to deviate significantly from spherical
symmetry. Note that the beam profiles used in the CO observations are
larger than $6''$, and thus measure the temperature, density and
velocity averaged in all directions in the regions where the lines are
mainly formed.  Hence, the main stellar and mass-loss properties can
be inferred from spherically symmetric models in fair
approximation. In reality, these average properties are formed
from a superposition of all these structures in this region.

\subsection{Comparison with other studies}
\label{comparison} 
In Sect.~\ref{masslossparameters}, we compare our results with
mass-loss parameters obtained in previous studies also tracing the
mass-loss variability of \object{WX Psc}.  In Sect.~\ref{scattered},
we focus on the $V$, $J$, $H$, and $K^{\prime}$ images of
\citet{Mauron2006A&A...452..257M} and
\citet{Hofmann2001AA...379..529H} probing both the dust-scattered
stellar and galactic light and the thermal dust emission.

\subsubsection{Mass-loss parameters} 
   \label{masslossparameters}

\begin{figure}[!thp]
  \includegraphics[height=0.5\textwidth,angle=90]{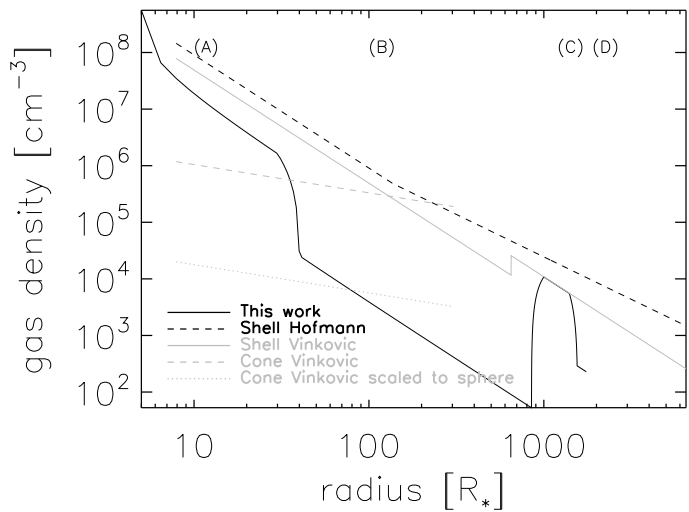}
  \caption{Number densities as a function of radial distance from the
    star as deduced by different authors. Full black line: spherical CSE
    of this study; dashed black line: spherical CSE of
    \citet{Hofmann2001AA...379..529H}; full grey line: spherical CSE of
    \citet{Vinkovic2004MNRAS.352..852V}; dashed grey line: bipolar cone
    of \citet{Vinkovic2004MNRAS.352..852V}; dotted grey line: number
    density of bipolar cone of \citet{Vinkovic2004MNRAS.352..852V} 
    scaled to a sphere.}
  \label{Mdot_comparison}
\end{figure}

\citet{Hofmann2001AA...379..529H} modeled the SED between 1\,$\mu$m
and 1\,mm, together with the visibility functions at 1.24~($J$ band),
1.65~($H$), 2.12 ($K^{\prime}$) and 11\,$\mu$m using {\sc dusty}
\citep{DUSTY}.  The visibility data provide information on the
innermost regions of the circumstellar environment, probing the
scattering properties of the dust (at $J$) and thermal emission
properties of the hot and warm dust (at $H$, $K$ and 11~$\mu$m). Assuming
a spherical outflow, they found that both the near-IR visibilities as
well as the mid-IR photometric points up to 100\,$\mu$m could be well
reproduced by a two-component outflow in which dust condenses at
900\,K. The most recent mass loss, characterized by a constant
velocity outflow (i.e.\ $\rho \propto r^{-2}$), is $\Mdot\,=\, 1.3 -
2.1 \times 10^{-5}$\,\Msun\,yr$^{-1}$. At distances beyond
135\,\Rstar\ the density decrease is shallower, matching $\rho \sim
r^{-1.5}$ (see the dashed black line in
Fig.~\ref{Mdot_comparison}). At wavelengths beyond about 400\,$\mu$m
this model underestimates the observed flux.

The $J$-band data reveal an asymmetry, and the $H$-band visibility
shows a puzzling drop in the $H$-band at spatial frequencies $\ga 14$ 
cycles per arcsec, corresponding to structure smaller than the 
condensation radius. The spherical model of 
\citet{Hofmann2001AA...379..529H} can not address these observations.
The above issues were taken up by \citet{Vinkovic2004MNRAS.352..852V}.
They explain the image asymmetries as originating from about 
10$^{-4}$\,\Msun\ of wind material that is swept-up
in an elongated cocoon whose expansion is driven by
bipolar jets and which extends out to a radial distance of
${\sim}$1100\,AU (or 300\,\Rstar), with an opening angle of 
${\sim}$30$^{\circ}$. The two cocoons have a characteristic lifetime
that is $\la 200$\,yr. To explain the mid-IR spectrum they 
invoke a spherical envelope (in which the cocoons are embedded)
containing ${\sim}$0.13\,\Msun\ of material extending out to 
${\sim}$35\arcsec\ (or $\sim$8\,000\,\Rstar). The size of the envelope
corresponds to a dynamical flow time of 5\,500 yr. Over 
time the mass loss rate has decreased. While $\Mdot \sim 2 \times 
10^{-5}$\,\Msun\,yr$^{-1}$ until 550\,yr ago, the most recent mass-loss 
is estimated to be only ${\sim}$9$\times 10^{-6}$\,\Msun\,yr$^{-1}$.
Two-dimensional radiative transfer calculations demonstrate that this 
model can explain both the $J$-band asymmetry, being dominated by 
scattered light escaping through the cone, and the spherical symmetry 
seen in the $H$ and $K^{\prime}$-band images, being due to predominantly 
thermal dust emission. This model also fits the general structure of 
the SED quite well, but has some problems in explaining the silicate 
feature around $10$~$\mu$m and the photometric data at $\lambda > 
100\,\mu$m.

\begin{figure*}[!thp]
  \sidecaption
  \includegraphics[height=12cm,angle=90]{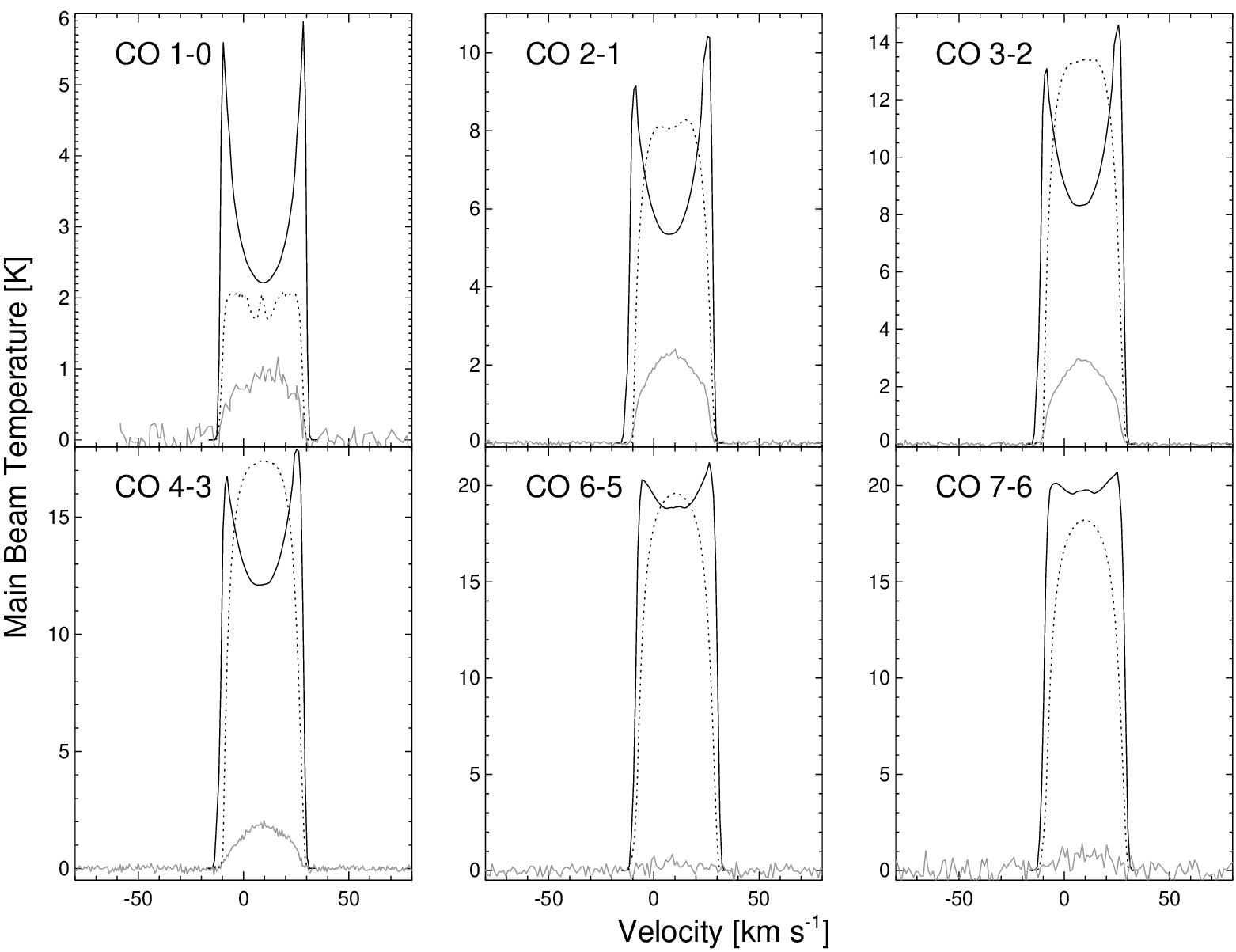}
  \caption{
    \label{WXPsc_model_others} CO rotational line profiles of
    \object{WX Psc} \citep[full grey line;][]{Nyman1992AAS...93..121N,
      Kemper2003AA...407..609K} compared with model predictions based on
    the parameters as proposed by \citet{Hofmann2001AA...379..529H}
    (dotted black line) and \citet{Vinkovic2004MNRAS.352..852V} (full
    black line).  }
\end{figure*}

In general, the mass-loss rate in the spherical envelope derived by
\citet{Hofmann2001AA...379..529H} and
\citet{Vinkovic2004MNRAS.352..852V} is much larger than what is
implied by the CO lines (see Fig.~\ref{Mdot_comparison}) and is not
consistent with the observed CO rotational line profiles (see
Fig.~\ref{WXPsc_model_others}). The CO (1$-$0) through (7$-$6) lines are
a sensitive probe of a geometrically extended part of the CSE, save
for the innermost region. This again may indicate that part of the
dust does not participate in driving the wind.

\subsubsection{Dust scattered light}\label{scattered}

Besides the work by \citet{Hofmann2001AA...379..529H} and
\citet{Vinkovic2004MNRAS.352..852V}, HST observations by
\citet{Mauron2006A&A...452..257M} also trace mass-loss variability in
the ambient environment of \object{WX Psc}. The HST-image shows part
of a faint circular arc at ${\sim}$0.8\arcsec (or ${\sim}$180\,\Rstar)
from the center. We performed additional simulations introducing a
third shell, positioned at this location (see
Sect.~\ref{sec:modell-co-rotat}).  These simulations demonstrate that
it may be possible that a faint arc is present with a mass-loss rate
$\la 2.5 \times 10^{-7}$\,\Msun\,yr$^{-1}$, although the
log-likelihood of this more sophisticated model is not significantly
better than for the simpler two-shell model.  The presently available
rotational CO line profiles can hence not be used to draw any
conclusions on the presence of this faint arc.

\citet{Mauron2006A&A...452..257M} have obtained the $B$ and
  $V$-band scattered light profiles of \object{WX Psc}. The scattered light is
  predominantly due to external illumination of the shell. These
  images show a relatively smooth decrease of intensity as a function
  of distance and at face-value do not seem to necessitate the
  invocation of mass-loss variability to explain them. However, the
  observed scattered light profile is very sensitive to the location
  of the external illumination sources and the phase function of
  scatterers. Formally the expression that these authors have used to
  fit to the observed profiles under the assumption of uniform
  illuminations is incorrect. The expression that they apply concerns
  the scattered part of the observed intensity, while under uniform
  illumination this part exactly compensates the intensity which is
  scattered out of the beam. It is at present unclear how a variable
  mass-loss history would translate into an observed intensity profile
  on the sky as a function of these parameters. It is obvious that a
  forward peaked scattering function will tend to de-emphasise any
  mass-loss variations at large distances. But whether this effect is
  sufficient to explain the smooth brightness distribution remains to
  be seen. A proper model for the light scattering in the CSE of
  \object{WX Psc}, taking into account its position in the galaxy,
  should be constructed. This, however, is clearly beyond the scope of
  this paper.

\subsubsection{Conclusion}

  To conclude, we use constraints from the dust and the CO molecules
  in the shell to determine the density profile within the few to few
  thousand stellar radii regime. The tracers we use are sensitive to
  these scales in a series of overlapping distance steps. The real
  power of our modeling comes from using \emph{both} the spectral
  information and the absolute strength of the observed quantities,
  be-it the molecular lines or the dust excess. These are all
  explained to satisfactory detail by the model presented here. The
  limitations of the method are also clear. While this yields good
  estimates for the average density as a function of distance, neither
  the large-beam observations nor the spherical model have the power
  to draw conclusion about small-scale ($\Delta R/R \approx 0.5)$
  structures or asymmetries.

\subsection{Mass-loss variability} 
\label{variability} 
As already noted in Sect.~\ref{spherical},
\citet{Mauron2006A&A...452..257M} concluded that \object{WX Psc}
resembles \object{IRC\,+10216} where approximate circular symmetry and
discrete shells exist in the extended envelope. In case of
\object{IRC\,+10216}, the shell spacing varies between 200 and
800\,yr, although intervals as short as 40\,yr are seen close to the
star \citep{Mauron2000A&A...359..707M}. The time spacing between shell
(A) and (C) in \object{WX Psc} is $\sim$800\,yr. The total envelope
mass is estimated to be 0.03\,\Msun, of which $\sim$95\,\% is located
in region (C). Assuming \object{WX Psc} to be a typical (solar-type)
AGB star losing some 0.5\,\Msun\ during its AGB evolution lasting some
100\,000\,yr, \object{WX Psc} has already lost $\sim$6\,\% of the
total mass that it will shed in its AGB-phase in a period of 600\,yr
(region C). Such a strong ejection event may indicate that \object{WX
Psc} is currently in its superwind phase. Assuming that a typical
oxygen-rich AGB star expels 0.1\,\Msun\ during the last 1000\,yr that
it experiences a superwind mass-loss phase
\citep{Marigo2007astro.ph..3139M}, few such high mass-loss phases may
occur during the superwind phase of \object{WX Psc}.

Various models have been proposed for the origin of discrete shells in
CSEs. One type of models is based on the effects of a binary companion
\citep[see, e.g.,][]{Harpaz1997ApJ...487..809H,
  Mastrodemos1999ApJ...523..357M}. A different scenario is invoked by
\citet{Simis2001A&A...371..205S}, who suggested that such shells are
formed by a hydrodynamical oscillation due to instabilities in the
gas-dust coupling in the CSE while the star is on the AGB.
\citet{Soker2000ApJ...540..436S} invoked another mechanism, being a
solar-like magnetic activity cycle in the progenitor AGB star.
\citet{Soker2006NewA...11..396S} raised the possibility that
semi-periodic oscillations in the photospheric molecular opacity may
be another candidate mechanism for the formation of multiple
semi-periodic arcs.

\section{Conclusions}
\label{conclusions}
In this paper, we have analyzed the rotational CO (1$-$0) through (7$-$6)
line profiles of \object{WX Psc} \emph{in combination} with the dust
characteristics in the full SED from 0.7 to 1300\,$\mu$m. This
combined analysis yields strong constraints on the density and
temperature structure over a geometrically large extent of the
circumstellar envelope, save for the innermost region.  Both CO line
fitting and dust modeling evidence strong mass-loss modulations during
the last 1700\,yr. Particularly, \object{WX Psc} underwent a high
mass-loss phase (\Mdot $\sim 5 \, 10^{-5}$\,\Msun/yr) lasting some
$600$\,yr and ending $\sim$800\,yr ago.  This period of high mass loss
was followed by a long period  of low mass loss (\Mdot
$\sim 5 \, 10^{-8}$\,\Msun/yr). The current mass loss is estimated to
be $\sim 6 \, 10^{-6}$\,\Msun/yr. We want to emphasize that the time
spacing of these mass loss modulations are independently well
constrained by both the molecular gas and dust modeling. The
uncertainty on the mass-loss rate in the different phases is estimated
to be a factor of a few, mainly caused by the uncertainty on the
dust-to-gas ratio and the CO/H$_2$ ratio, and by the assumption of a
homogeneous spherically symmetric envelope. Most importantly, the
mass-loss history that we derive is significantly different from a
constant mass-loss rate model, and moreover implies that \object{WX
Psc} can have at maximum in the order of a few high mass-loss phases
during its final superwind evolution on the AGB.

\begin{acknowledgements}
  LD and SD acknowledge financial support from the Fund for Scientific
  Research - Flanders (Belgium), SH acknowledges financial support
  from the Interuniversity Attraction Pole of the Belgian Federal
  Science Policy P5/36. The computations for this research have been
  done on the VIC HPC Cluster of the KULeuven. We are grateful to the
  LUDIT HPC team for their support. We would like to thank Remo
  Tilanus (JCMT) for his support during the observations and reduction of
  the data. 
\end{acknowledgements} 


\end{document}